\documentclass[12pt]{iopart}

\usepackage[utf8]{inputenc}
\usepackage[T1]{fontenc}
\usepackage{graphicx}
\usepackage{scrextend}
\usepackage{units}
\usepackage{cite}
\usepackage{units}

\newcommand{\sfig}[2]{\mbox{figure \ref{#1}(#2)}}
\newcommand{\tab}[1]{\mbox{table \ref{#1}}}
\newcommand{\eq}[1]{\mbox{equation (\ref{#1})}}
\newcommand{\iu}{{i\mkern1mu}}
\newcommand{\ensave}[1]{\langle\langle#1\rangle\rangle}
\newcommand{\abs}[1]{\left|#1\right|}

\begin{document}
\title[]{Limitations on the indistinguishability of photons from remote solid state sources}
\author{Benjamin Kambs and Christoph Becher}
\address{Fachrichtung Physik, Universität des Saarlandes, Campus E 2.6, 66123 Saarbrücken, Germany}
\ead{christoph.becher@physik.uni-saarland.de}
\vspace{10pt}
\begin{indented}
\item[]June 2018
\end{indented}
\begin{abstract}
In the present work, we derive a formalism that can be used to predict and interpret the time structure and achievable visibilities for two-photon interference (TPI) experiments using photons from two separate sources. The treatment particularly addresses photons stemming from solid state quantum emitters, which are often subject to pure dephasing (PD) and spectral diffusion (SD). Therefore, it includes the impact of phase- and emission frequency-jitter besides the influence of differing radiative lifetimes and a relative spectral detuning. While the treatment is mainly aimed at interference experiments after Hong-Ou-Mandel (HOM), we additionally offer generalized equations that are applicable to arbitrary linear optical gates, which rely on TPI.
\end{abstract}
%
%
\vspace{2pc}
\noindent{\it Keywords}: quantum communication, quantum computing, solid state single photon sources, two-photon interference, pure dephasing, spectral diffusion
\section{Introduction}
Two-photon interference between indistinguishable photons is at the heart of key quantum technologies, such as linear optical quantum computing \cite{Knill2001, Kok2007, Aaronson2013, Tillmann2013} and entanglement distribution within quantum repeater networks \cite{Briegel1998,Duan2001,Sangouard2011,Takeoka2014,Pirandola2017}. Therefore, various types of quantum emitters have been investigated as possible sources of indistinguishable single photons over the past decades. Most prominent among them are spontaneous parametric down conversion sources \cite{Hong1987, Kaltenbaek2006}, trapped ions \cite{Maunz2007} and atoms \cite{Legero2004, Beugnon2006} as well as single emitters in solid state host materials like semiconductor quantum dots \cite{Santori2002, Somaschi2016}, defect centers in diamond \cite{Bernien2012, Sipahigil2014} or single molecules in crystalline host matrices \cite{Kiraz2005, Ahtee2009}. The degree of indistinguishability is commonly assessed by interferometric measurements following the pioneering Hong-Ou-Mandel experiment \cite{Hong1987}. For best interference visibilities the interacting photons need to be in the same quantum state. In particular poor spectral, temporal or spatial mode overlaps as well as non-matching polarization states lead to a strong corruption of the TPI signature. Well engineered and controlled state-of-the-art sources provide nearly Fourier-transform limited single photons that achieve close to optimal interference visibilities between consecutively emitted photons from a single source \cite{Somaschi2016, Wang2016}.\par
For TPI schemes used in systems of two or more distinct quantum nodes, it is mandatory that all participating emitters mutually meet the same conditions in order to accomplish best possible performances. In contrast to trapped ions and atoms, the spectral and temporal properties of solid state emitters strongly differ of one another due to a variety of interactions between the emitter and its environment \cite{Aharonovich2016}. First TPI experiments using photons stemming from two separate solid state sources employed techniques like strain-, temperature-, or electric field tuning to bring the emitters to a common bus wavelength and thus maximize their indistinguishability \cite{Patel2010, Bernien2012, Gold2014, Giesz2015, Reindl2017}. However, the achieved interference contrasts remain far behind corresponding values for experiments with photons from a single source and often even below the classical limit of \unit[50]{\%} for two independent coherent light sources \cite{Mandel1983, Ou1988}. Besides non-matching radiative lifetimes, the main culprit limiting the indistinguishability is the solid state host material itself: interactions with phonons as well as electrostatic fluctuations in the vicinity of the emitter destabilize its emission frequency. These perturbations either lead to pure dephasing (short interaction-time) or spectral diffusion (long interaction-time) resulting in homogeneous or inhomogeneous broadening of the emission line, respectively \cite{Kubo1969, Tokmakoff2014}. The timescale, at which SD occurs, is typically in the same order of or longer than the repetition time of the excitation cycles \cite{Thoma2016}. Therefore, even though the emission wavelength wanders, there is still a remaining correlation between the carrier frequencies of consecutively emitted photons. In case of single source experiments this correlation limits the loss of indistinguishability. Remote solid state emitters on the other hand diffuse independently, i.e. no such correlation exists between the photons under investigation. Accordingly, instead of the SD rate, the overall distribution of emission wavelengths becomes the relevant factor influencing the observed interference contrast.\par
In order to prepare available emitters for remote TPI applications in quantum devices, a quantitative understanding of their limits is essential. To this point most experimental data were supported by theoretical models based on \cite{Legero2003, Kiraz2004}. Both formalisms investigate the HOM effect in the time domain, but have complementary scopes: The equations presented in \cite{Legero2003} are derived for two arbitrary input fields based on the underlying single photon wave functions. This allows to predict experiments with two dissimilar single photons at the cost of an elaborate integration step. In contrast, \cite{Kiraz2004} is restricted to two identical photons, but only requires knowledge of their coherence functions. Applications of both formalisms include simulations \cite{Flagg2010, Patel2010, Bernien2012} as well as analytical treatments of either PD \cite{Bylander2003, Giesz2015, Reindl2017} or SD \cite{Gold2014}. However, those models are typically tailored to describe the data at hand and thus can only be applied to a limited range of experimental conditions. The present work offers an analytical treatment of correlation measurements on systems that rely on TPI between single photons as emitted by two remote solid state sources. The approach simultaneously aims at generality and easily applicable results. On that account, we first extend the well-established formalism for TPI at a single beam-splitter \cite{Legero2003} to the scattering of two photons at an arbitrary linear optical gate. The resulting equation is then integrated for a set of emitters with differing radiative lifetimes, relative detuning of their carrier frequencies and under the influence of PD as well as SD. Moreover, a treatment of technical imperfections such as synchronization and polarization mismatch at the gate inputs and imperfect optical components of the gate itself is offered in the supplement. The results are verified by comparison to various experiments and accompanying models as presented in literature. To further demonstrate the capabilities of our formalism, we apply it in order to emulate the entanglement generation by means of a CNOT gate \cite{Ralph2002,OBrien2003}. Eventually, we use the results to evaluate the achievable Bell-state fidelity, if the gate was operated with state-of-the-art quantum emitters.
\section{Derivation}
\begin{figure}
	\centering
	\includegraphics[width= \columnwidth]{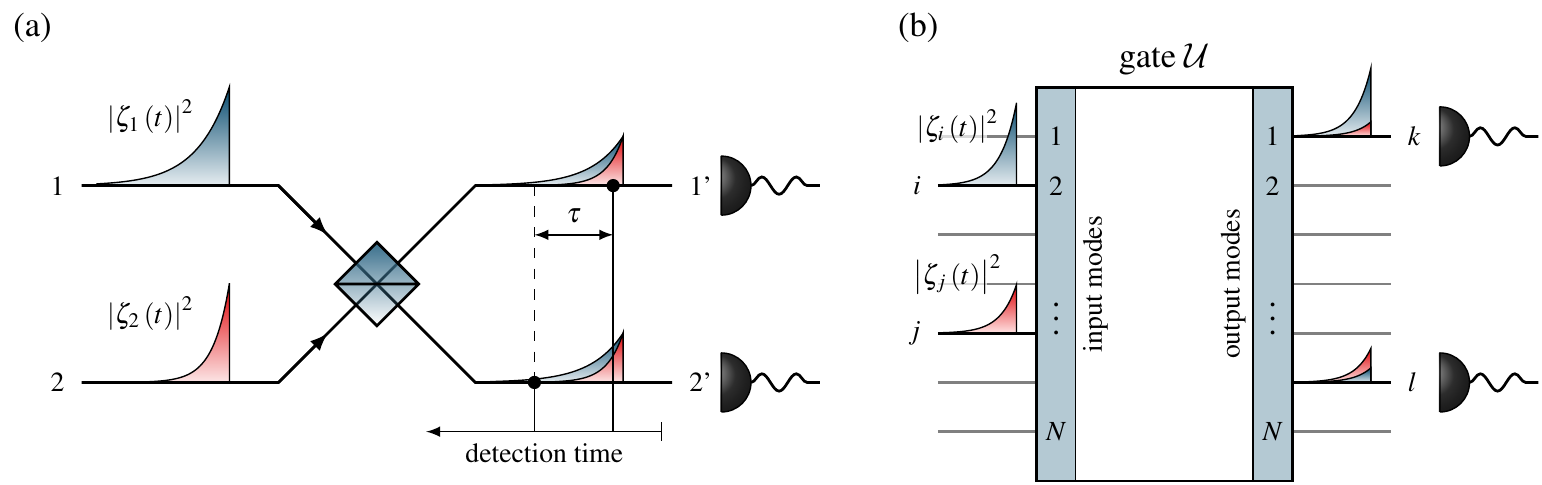}%
	\caption{(a) Schematic Hong-Ou-Mandel experiment: Two photons, here represented by their field modes $\left|\zeta_{1,2}\left(t\right)\right|^2$, enter a symmetric beam splitter via the inputs 1 and 2. If both photons are identical, they will never leave through separate outputs. Experimentally, this is verified by a vanishing cross-correlation of detection events between the detected signals of both outputs around $\tau = 0$. (b) Hong-Ou-Mandel experiment generalized to an arbitrary gate $\mathcal{U}$ with $N$ inputs and $N$ outputs as considered in the present work: Two photons entering via the inputs $i$ and $j$ are scattered by the gate. Coincidence events between the outputs $k$ and $l$ are measured.}%
	\label{fig1}
\end{figure}

The fundamental HOM-effect is described with two identical single photons entering a symmetric beam splitter through two different input arms 1 and 2 as schematically depicted in \sfig{fig1}{a}. With $a_1^\dagger$ and $a_2^\dagger$ being the creation operators for photons in these channels, the initial situation is represented by the Fock-state $a_1^\dagger a_2^\dagger\left|0\right.\rangle = \left|1_11_2\right.\rangle$. The unitary transformation relations of the beam splitter to the output channels $1'$ and $2'$ using their creation operators $a{'}_1^\dagger$ and $a{'}_2^\dagger$, respectively, are given as
\begin{equation}
a_1^\dagger = \frac{1}{\sqrt{2}} \left(a{'}_1^\dagger +a{'}_2^\dagger\right) \quad \mbox{and} \quad
a_2^\dagger = \frac{1}{\sqrt{2}} \left(a{'}_1^\dagger -a{'}_2^\dagger\right).
\label{bstransform}
\end{equation}
Applying these equations, it is straightforward to show that the input state is projected to $\left(\left|2_{1'}0_{2'}\right.\rangle-\left|0_{1'}2_{2'}\right.\rangle\right)/\sqrt{2}$, i.e. both photons always leave through the same output port. Commonly, this bunching signature is attributed to the bosonic nature of photons. Although this treatment allows to approach the physics of indistinguishable particles using textbook knowledge only, it fails to describe all realistic situations in which photons evolve in more complex field modes. To cover the interaction of arbitrary input fields, the transformations (\ref{bstransform}) can be expressed in terms of their field operators
\begin{equation}
\hat{E}^+_k\left(t\right) = \zeta_k \left( t\right) \hat{a}_k \quad \mbox{and} \quad \hat{E}^-_k\left(t\right) = \zeta^*_k \left( t\right) \hat{a}^\dagger_k,
\label{fieldoperator}
\end{equation}	
with $k$ being any of the two inputs and $\zeta_k\left(t\right)$ the normalized field modes of the photons. In the course of this approach it was shown that the probability to obtain detection events at time $t_0$ and $t_0+\tau$ at channels $1'$ and $2'$, respectively, reads \cite{Legero2003}
\begin{equation}
P_{joint} \left( t_0, \tau \right) = \frac{1}{4}\left|\zeta_1\left(t_0+\tau\right)\zeta_2\left(t_0\right)-\zeta_2\left(t_0+\tau\right)\zeta_1\left(t_0\right)\right|^2.
\label{pjointlegero}
\end{equation}
This relation enables to compute the second order cross-correlation function $g^{\left(2\right)}\left(\tau\right)$ via
\begin{equation}
g^{\left(2\right)}\left(\tau\right) = \int_{-\infty}^{+\infty}{P_{joint} \left( t_0,\,\tau \right)\,\mbox{d}t_0},
\label{g2define}
\end{equation}
which is the quantity of interest in HOM-experiments, as no correlated events are detected, if both photons leave the beam splitter via the same output. Thus, the number of correlated events around $\tau = 0$ is reduced compared to classical light depending on the degree of indistinguishability.\par
In the following section an expression for a general linear optical gate equivalent to \eq{pjointlegero} is derived. Assuming two single photons at two arbitrary inputs of the gate, integral (\ref{g2define}) is then solved to yield the measured cross-correlation function between any two output ports. Eventually, a final integration step with respect to the time lag $\tau$ yields the overall coincidence probability.
\subsection{Joint detection probability for linear optical gates}
Linear optical gates are networks of passive linear optical components that accept photons propagating within $N$ input modes and map them to $N$ output modes (compare \sfig{fig1}{b}). Any linear gate is represented by a unitary $N\times N$ matrix $\mathcal{U}$, which can be used to find the field operator $\hat{E}^{'\pm}_k\left(t\right)$ of output mode $k$ depending on all input fields $\hat{E}^{\pm}_l\left(t\right)$ according to
\begin{equation}
\hat{E}^{'+}_k\left(t\right) = \sum_{l}{\mathcal{U}_{kl} \hat{E}^+_l\left(t\right)} \quad \mbox{and} \quad \hat{E}^{'-}_k\left(t\right) = \sum_{l}{\mathcal{U}^*_{kl} \hat{E}^-_l\left(t\right)},
\label{transfieldoperator}
\end{equation}
where $\mathcal{U}_{kl}$ are the elements of $\mathcal{U}$. Given two photons at distinct inputs $i$ and $j$, the joint probability $P_{joint} \left(t_0,\tau\right)$ for two detection events at the outputs $k$ and $l$ at times $t_0$ and $t_0 + \tau$, respectively, reads in agreement with \cite{Legero2003}\footnote{Note that a rigorous notation should include the dependency of $P_{joint} \left(t_0,\tau\right)$ on all participating modes $i$, $j$, $k$ and $l$. We omit these dependencies in favor of legibility.}
\begin{equation}
P_{joint} \left(t_0,\tau\right) = \langle 1_i1_j | \hat{E}^{'-}_k\left(t_0\right) \hat{E}^{'-}_l\left(t_0+\tau\right) \hat{E}^{'+}_l\left(t_0+\tau\right) \hat{E}^{'+}_k\left(t_0\right)  | 1_i1_j \rangle.
\label{pjoint}
\end{equation}
With help of the transformation relation (\ref{transfieldoperator}) and the definition of field operators (\ref{fieldoperator}), $P_{joint} \left(t_0,\tau\right)$ can be expanded to
\begin{eqnarray}
P_{joint} \left(t_0,\tau\right) = \sum_{r,s,u,v}&\mathcal{U}^*_{kr}\mathcal{U}^*_{ls}\mathcal{U}_{lu}\mathcal{U}_{kv}\zeta^*_r \left( t_0\right)\zeta^*_s \left( t_0+\tau\right)\zeta_u \left( t_0+\tau\right)\zeta_v \left( t_0\right)
\nonumber \\
&\times\langle 1_i1_j  | \hat{a}^\dagger_r\hat{a}^\dagger_s \hat{a}_u\hat{a}_v | 1_i1_j  \rangle.
\end{eqnarray}
The summation over $r,s,u,v$ accounts for all contributing input modes. Accordingly, all of these indices must be either $i$ or $j$, leaving only 16 summands. Using $\hat{a}| 0  \rangle = 0$ it is straightforward to show that just 4 of those are nonzero. Eventually, they can be factorized to
\begin{equation}
P_{joint} \left(t_0,\tau\right) = \left| \mathcal{U}_{li}\mathcal{U}_{kj} \zeta_i \left( t_0+\tau\right)\zeta_j \left( t_0\right) + \mathcal{U}_{lj}\mathcal{U}_{ki} \zeta_j \left( t_0+\tau\right)\zeta_i \left( t_0\right)\right|^2.
\label{pjointgeneral}
\end{equation}
This expression is the generalized modification of \eq{pjointlegero}. It gives access to all quantities of interest that used to be evaluated for single beam splitter HOM experiments without any additional complexity, if only the gate matrix $\mathcal{U}$ is known.
\subsection{Cross-correlation function}
While \eq{pjointgeneral} is applicable to arbitrary input fields, we focus on single photons as emitted by solid state emitters in the scope of the present work. An ideal quantum emitter can be treated as a two-level system. Subsequent to its excitation it relaxes to the ground state via spontaneous emission of a single photon. Accordingly, the single photon wave functions arriving at the gate inputs $i$ and $j$ can be described by a one sided exponential decay as
\begin{equation}
\zeta_{i,j} \left(t\right) =  \frac{1}{\sqrt{\tau_{i,j}}}\mbox{H}\left(t\right)\cdot \exp\left\{-\nicefrac{t}{2\tau_{i,j}} - \iu \left[2\pi\nu_{i,j}t + \varphi_{i,j} \left(t\right)\right]\right\}
\label{wfansatz}
\end{equation}
where $\tau_{i,j}$ is the excited state radiative lifetime and $\nu_{i,j}$ the carrier frequency of the emitted photon. The Heaviside-function $\mbox{H} \left(t\right)$ takes into account that no photon can be emitted prior to the excitation step. The time-dependent phase $\varphi_{i,j} \left( t \right)$ is introduced to model PD. Furthermore, we assume that $\nu_{i,j}$ remains constant for the duration of a single photon, which is valid in the limit of slow SD compared to $\tau_{i,j}$. The wave functions are normalized, i.e. we have
\begin{equation}
\int_{-\infty}^{+\infty}{\left|\zeta_{i,j} \left(t\right)\right|^2\,\mbox{d}t} = 1.
\end{equation}
Using definition (\ref{wfansatz}), it follows that the joint photon-detection probability (\ref{pjointgeneral}) is given by
\begin{equation}
P_{joint} \left( t_0,\,\tau \right) = \frac{1}{\tau_i\tau_j}\cdot f\left(t_0,\,\tau\right) \cdot g\left(t_0,\,\tau\right)
\label{pjoint2}
\end{equation}
with the terms
\begin{eqnarray}
f\left(t_0,\,\tau\right) 	= &   \left|\mathcal{U}_{li}\right|^2\left|\mathcal{U}_{kj}\right|^2\cdot\exp\left(-\nicefrac{\tau}{\tau_i}\right) + \left|\mathcal{U}_{lj}\right|^2\left|\mathcal{U}_{ki}\right|^2\cdot\exp\left(-\nicefrac{\tau}{\tau_j}\right)
\nonumber \\
&+ \left|\mathcal{U}_{li}\mathcal{U}_{kj}\mathcal{U}^*_{ki}\mathcal{U}^*_{lj}\right|\cdot \exp\left(-\nicefrac{\tau}{2T_+}\right)\cdot h\left(t_0,\,\tau\right),
\nonumber \\
g\left(t_0,\,\tau\right) 	= &\mbox{H} \left(t_0\right)\mbox{H} \left(t_0+\tau\right)\cdot \exp\left(-\nicefrac{t_0}{T_+}\right), \quad \mbox{and}
\nonumber \\
h\left(t_0,\,\tau\right) 	= & 2\cos{\left(2\pi\Delta\nu\tau + \Delta\varphi_i - \Delta\varphi_j + \Phi_\mathcal{U}\right)}.
\label{pjs}
\end{eqnarray}
Also, a number of abbreviations were used: $\Phi_\mathcal{U} = \arg{\left(\mathcal{U}_{li}\mathcal{U}_{kj}\mathcal{U}^*_{ki}\mathcal{U}^*_{lj}\right)}$ for the phase introduced by the gate, \mbox{$\Delta\nu = \nu_i - \nu_j$} being the momentary frequency displacement between the emitters, the phase differences \mbox{$\Delta\varphi_{i,j} = \varphi_{i,j}\left(t_0 + \tau\right) - \varphi_{i,j}\left(t_0\right)$}, as well as \mbox{$1/T_+ = 1/\tau_i + 1/\tau_j$}. As PD and SD lead to random fluctuations of $\varphi_{i,j}$ and $\nu_{i,j}$ and correspondingly to an unknown time-dependency of the phase-sensitive term $h\left(t_0,\,\tau\right)$, integral (\ref{g2define}) cannot be directly solved. However, HOM measurements are typically integrated over a long time and yield a cross-correlation function $\mathcal{G}^{\left(2\right)}\left(\tau\right)$, which is then averaged over all occurring $\Delta\nu$ and $\Delta\varphi_{i,j}$ according to
\begin{equation}
\mathcal{G}^{\left(2\right)}\left(\tau\right)=\int_{-\infty}^{+\infty}{\ensave{P_{joint} \left(t_0,\tau\right)}\,\mbox{d}t_0},
\label{g2ave}
\end{equation}
where $\ensave{\,\cdot\,}$ denotes the statistical averaging. Considering that PD and SD act independently on the emitters, the known treatments of both individual effects \cite{Bylander2003,Gold2014} can be used to evaluate the average, leading to
\begin{equation}
\ensave{h\left(t_0,\,\tau\right)} = 2 \exp\left[-\left(\Gamma^*_i + \Gamma^*_j \right) \left|\tau\right|-2\pi^2\Sigma^2\tau^2\right]\cdot\cos{\left(2\pi\delta\nu\tau-\Phi_\mathcal{U}\right)}.
\label{haverage}
\end{equation}
The solution includes joint spectral properties of both emitters, namely the pure dephasing rates $\Gamma^*_{i,j}$, the inhomogeneous linewidth contributions given by their standard deviations $\sigma_{i,j}$ contained in $\Sigma^2 = \sigma_i^2+\sigma_j^2$, as well as the relative detuning of both emission lines $\delta\nu$. Note that $\sigma$ is only used for the purpose of a concise notation. A more common para\-meter characterizing the inhomogeneous broadening is the full width at half maximum $\sigma'$, which is connected to the standard deviation by $\sigma' = 2\sqrt{2\ln2} \cdot \sigma$. Details on how to evaluate the statistical average and how all relevant parameters are connected to first order correlation and spectral measurements of the single emitters can be found in the supplement. Using \eq{haverage}, we eventually obtain
\begin{eqnarray}
\fl \mathcal{G}^{\left(2\right)}\left(\tau\right) =& \mathcal{G}_0^{\left(2\right)} \left(\tau\right) + \mathcal{G}_{int}^{\left(2\right)} \left(\tau\right) &\mbox{with}
\nonumber \\
\fl \mathcal{G}_0^{\left(2\right)} \left( \tau \right) = &\frac{1}{\tau_i + \tau_j} \left\{ \left|\mathcal{U}_{li}\right|^2\left|\mathcal{U}_{kj}\right|^2\left[\mbox{H}\left(\tau \right)\cdot\exp\left(-\nicefrac{\tau}{\tau_i}\right)+\mbox{H}\left(-\tau \right)\cdot\exp\left(\nicefrac{\tau}{\tau_j}\right)\right]\right.
\nonumber \\
\fl &\phantom{\frac{1}{\tau_i + \tau_j} \left\{\right.}+ \left. \left|\mathcal{U}_{lj}\right|^2\left|\mathcal{U}_{ki}\right|^2\left[\mbox{H}\left(\tau \right)\cdot\exp\left(-\nicefrac{\tau}{\tau_j}\right)+\mbox{H}\left(-\tau\right)\cdot\exp\left(\nicefrac{\tau}{\tau_i}\right)\right]\right\} \quad &\mbox{and}
\nonumber \\
\fl \mathcal{G}_{int}^{\left(2\right)} \left( \tau \right) = & \frac{2\left|\mathcal{U}_{li}\mathcal{U}_{kj}\mathcal{U}^*_{ki}\mathcal{U}^*_{lj}\right|}{\tau_i + \tau_j}\cdot \exp\left(-\gamma\left|\tau\right|-2\pi^2\Sigma^2\tau^2\right)\cdot \cos{\left(2\pi\delta\nu\tau-\Phi_\mathcal{U}\right)}.
\label{g2solution}
\end{eqnarray}
The pure dephasing rates of both emitters are now included in the quantity $\gamma = \gamma_i + \gamma_j$ with $\gamma_{i,j} = 1/\left(2\tau_{i,j}\right) + \Gamma^*_{i,j}$.	Intermediate results for the integration over $t_0$ in \eq{g2ave} are extensive and offer no valuable insight. However, for the sake of completeness they are summarized in the supplement.\par
Equation (\ref{g2solution}) describes the time-structure of cross-correlation measurements between the outputs $k$ and $l$ of an arbitrary linear optical gate $\mathcal{U}$ for two photons given by the wave-function (\ref{wfansatz}) entering the gate via the inputs $i$ and $j$. Both photons can be subject to PD and SD and spectrally detuned with respect to each other. It is worth mentioning that the term $\cos{\left(2\pi\delta\nu\tau\right)}$ in $\mathcal{G}_{int}^{\left(2\right)}$ gives rise to the known quantum beats in TPI experiments \cite{Legero2004}. Also, with decreasing indistinguishability of both photons $\mathcal{G}_{int}^{\left(2\right)}$ converges towards zero. Consequently, $\mathcal{G}_0^{\left(2\right)}$ corresponds to the limit of $\mathcal{G}^{\left(2\right)}$ for entirely distinguishable photons.
\subsection{Overall coincidence probability}
The main figure of merit to quantify indistinguishability of two photons by means of a HOM experiment is the achieved interference visibility $V$. With $p_{coinc}$ being the average probability to measure a coincidence between the beam splitter outputs, the visibility can be evaluated by
\begin{equation}
V = 1- p_{coinc} / p_{coinc,0},
\label{visib}
\end{equation}
where $p_{coinc,0}$ denotes the classical limit of $p_{coinc}$ (=0.5 in case of a symmetric beam splitter). Accordingly, the visibility varies between $V=$ 0 for entirely distinguishable and $V=$ 1 for entirely indistinguishable photons. In a more general sense, $p_{coinc}$ contains information on the success probability of a given gate operation based on TPI and is often used to evaluate the fidelity of a desired output state \cite{James2001, White2007}. It can be accessed by integration over $\mathcal{G}^{\left(2\right)}\left(\tau\right)$ according to
\begin{eqnarray}
p_{coinc} & = \int_{-\infty}^{+\infty}{\mathcal{G}^{\left(2\right)}\left(\tau\right)\,\mbox{d}\tau}
\nonumber \\
& = \int_{-\infty}^{+\infty}{\mathcal{G}_0^{\left(2\right)}\left(\tau\right)\,\mbox{d}\tau} + \int_{-\infty}^{+\infty}{\mathcal{G}_{int}^{\left(2\right)}\left(\tau\right)\,\mbox{d}\tau}
\nonumber \\
& = p_{coinc,0} + p_{coinc,int}.
\label{pcoinc}
\end{eqnarray}
It is then straightforward to show that the first term merely yields
\begin{equation}
p_{coinc,0} = \int_{-\infty}^{+\infty}{\mathcal{G}_0^{\left(2\right)}\left(\tau\right)\,\mbox{d}\tau} = \left|\mathcal{U}_{li}\right|^2\left|\mathcal{U}_{kj}\right|^2 + \left|\mathcal{U}_{lj}\right|^2\left|\mathcal{U}_{ki}\right|^2.
\label{pzero}
\end{equation}
In order to solve the second integration over $\mathcal{G}_{int}^{\left(2\right)}\left(\tau\right)$, we first simplify by introducing the abbreviations $\alpha = 2\pi\Sigma$ and $\omega = 2\pi\delta\nu$. Also, we expand the $\cos$-function as $2\cos{\left(\omega\tau-\Phi_\mathcal{U}\right)} = \exp{\left(\iu\omega\tau-\iu\Phi_\mathcal{U}\right)} + c.c.$ and use the fact that the integrand is mirror-symmetric around $\tau = 0$, which leads to
\begin{equation}
\fl p_{coinc,int} = \frac{2\left|\mathcal{U}_{li}\mathcal{U}_{kj}\mathcal{U}^*_{ki}\mathcal{U}^*_{lj}\right|}{\tau_i + \tau_j}\cdot\cos{\left(\Phi_\mathcal{U}\right)}\cdot \underbrace{\int_{-\infty}^{+\infty}{\exp\left(-\gamma\left|\tau\right|-\frac{1}{2}\alpha^2\tau^2-\iu\omega\tau\right)\,\mbox{d}\tau}}_{=\sqrt{2\pi}\cdot\mathcal{F}\left[f\left(\tau\right)\cdot g\left(\tau\right)\right]\left(\omega\right)}.
\label{fourier}
\end{equation}
It can be seen that the term $\mathcal{F}\left[f\left(\tau\right)\cdot g\left(\tau\right)\right]\left(\omega\right)$ with $f\left(\tau\right) = \exp{\left(-\gamma\left|\tau\right|\right)}$ and $g\left(\tau\right) = \exp{\left(-\alpha^2\tau^2/2\right)}$ constitutes a unitary Fourier transform. Applying the convolution theorem, we then get
\begin{equation}
\fl p_{coinc,int} = \frac{2\left|\mathcal{U}_{li}\mathcal{U}_{kj}\mathcal{U}^*_{ki}\mathcal{U}^*_{lj}\right|}{\tau_i + \tau_j}\cdot\cos{\left(\Phi_\mathcal{U}\right)}\cdot \left(\sqrt{\frac{2}{\pi}}\frac{\gamma}{\gamma^2+\omega^2}\right)\ast\left(\frac{\exp\left(-\nicefrac{\omega^2}{2\alpha^2}\right)}{\alpha}\right),
\end{equation}
where the known Fourier transformations of an exponential decay and a Gaussian distribution were used and $\ast$ represents the convolution operator. The convolution of the Lorentzian and Gaussian function at hand defines a Voigt profile and cannot be analytically solved. However, it can be expressed in terms of the real part of the Faddeeva function $w\left(z\right)$ \cite{Abramowitz1964}, for which fast numerical implementations with high accuracy are available \cite{Poppe1990, Zaghloul2011}. With $z = \left(2\pi\delta\nu+\iu\gamma\right) / \left(2\pi\sqrt{2}\Sigma\right)$ this yields
\begin{equation}
p_{coinc,int} = 2\left|\mathcal{U}_{li}\mathcal{U}_{kj}\mathcal{U}^*_{ki}\mathcal{U}^*_{lj}\right|\cdot\cos{\left(\Phi_\mathcal{U}\right)}\cdot \frac{\mbox{Re}\left[w\left(z\right)\right]}{\sqrt{2\pi}\Sigma\left(\tau_i+\tau_j\right)}
\label{pint}
\end{equation}
The solutions (\ref{pzero}) and (\ref{pint}) can now be used to compute the desired coincidence probability $p_{coinc}$ according to \eq{pcoinc}
\begin{equation}
\fl p_{coinc} = \left|\mathcal{U}_{li}\right|^2\left|\mathcal{U}_{kj}\right|^2 + \left|\mathcal{U}_{lj}\right|^2\left|\mathcal{U}_{ki}\right|^2 + 2\left|\mathcal{U}_{li}\mathcal{U}_{kj}\mathcal{U}^*_{ki}\mathcal{U}^*_{lj}\right|\cdot\cos{\left(\Phi_\mathcal{U}\right)}\cdot \frac{\mbox{Re}\left[w\left(z\right)\right]}{\sqrt{2\pi}\Sigma\left(\tau_i+\tau_j\right)}.
\label{pcoincresult}
\end{equation}	
We see that $p_{coinc}$ follows a Voigt line shape as a function of the detuning $\delta\nu$. This result comes as no surprise, as it reflects the spectral properties of the emitters. The coincidence probability can be evaluated for arbitrary gate inputs and outputs as well as any set of two single photon wave-packets in the same manner as the cross-correlation function (\ref{g2solution}). More general expressions for $\mathcal{G}^{\left(2\right)}\left(\tau\right)$ and $p_{coinc}$ including the effects of asynchronous arrival and polarization mismatch between both photons at the gate inputs can be found in the supplement. 
\section{Discussion}
In the following we explore possible predictions that can be extracted from the presented formalism. First, we focus on HOM measurements, where $\mathcal{U}$ is nothing but a simple beam splitter matrix, and offer an expression for the TPI visibility. The result is used to assess our formalism by comparison to existing models and experimental data. Following \cite{Bylander2003}, the remote HOM visibility is often quantified using the coherence time of the emitters. We briefly review this relation, if the emitters are affected by both PD and SD. Eventually, to demonstrate the applicability to more complex systems, we investigate the entanglement generation performed by a CNOT gate. Here, the gate matrix $\mathcal{U}$ includes state preparation, the CNOT gate itself, and the final state tomography corresponding to a system of allover 11 linear optical components.
\subsection{HOM measurements}
\begin{figure}
	\centering
	\includegraphics[width= \columnwidth]{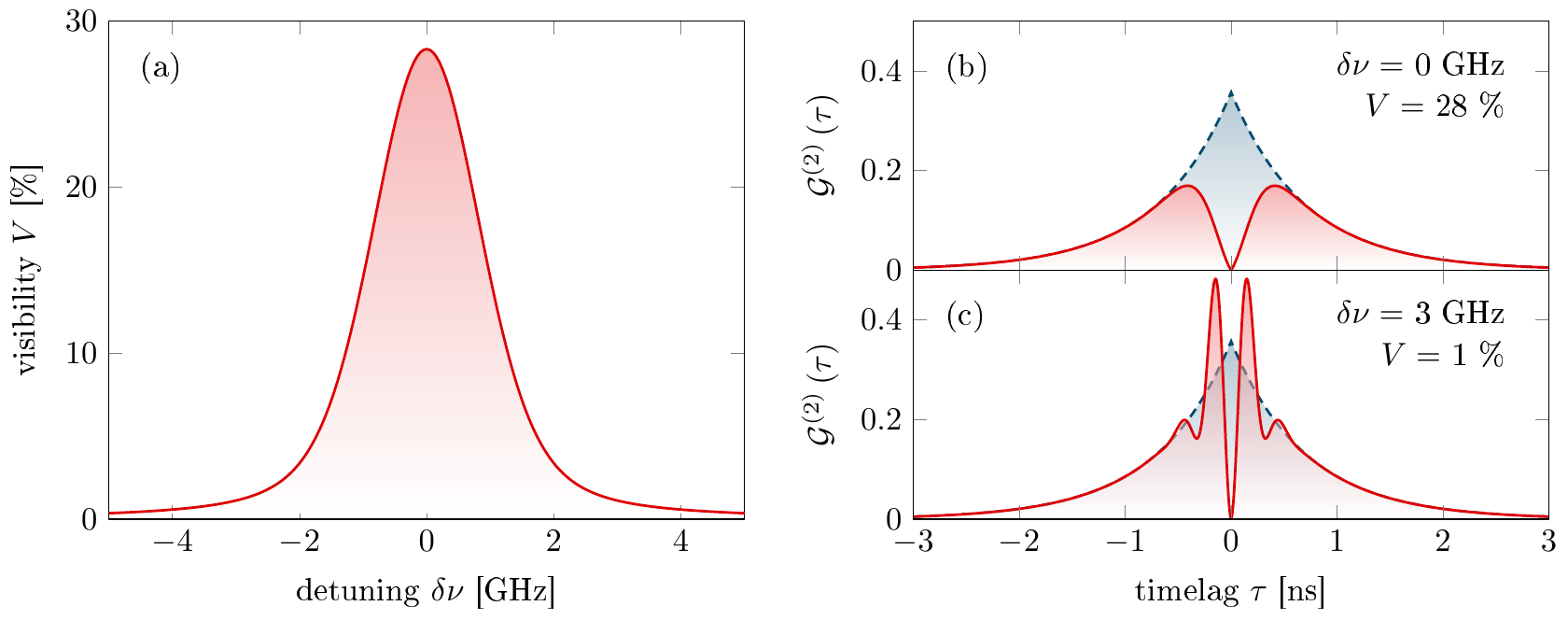}%
	\caption{(a) Visibility as a function of detuning for a set of emitters with \mbox{$\tau_1 =$ \unit[700]{ps}}, \mbox{$\tau_2 =$ \unit[650]{ps}}, \mbox{$\sigma'_1$ = \unit[1.4]{GHz}}, \mbox{$\sigma'_2$ = \unit[0.8]{GHz}}, \mbox{$\Gamma^*_1$ = \unit[600]{MHz}}, \mbox{$\Gamma^*_2$ = \unit[300]{MHz}}. The corresponding cross-correlation functions $\mathcal{G}^{\left(2\right)}\left(\tau\right)$ at relative detunings of $\delta\nu = $ \unit[0]{GHz} and $\delta\nu = $ \unit[3]{GHz} are shown in (b) and (c), respectively (solid red). The blue dashed curves show the classical limit for completely distinguishable photons.}
	\label{fig2}
\end{figure}

The coalescence of two photons at a single beam splitter corresponding to \sfig{fig1}{a} constitutes the most simple experiment exhibiting TPI. Accordingly, HOM measurements are performed by default to quantify the indistinguishability of single photons. The unitary matrix $\mathcal{U}_{BS}$ describing the action of a beam splitter is given by
\begin{equation}
\mathcal{U}_{BS} =
\left(\begin{array}{cc}
\mathcal{U}_{11} & \mathcal{U}_{12} \\
\mathcal{U}_{21} & \mathcal{U}_{22}
\end{array}\right) =
\left(\begin{array}{cc}
\sqrt{R} & \sqrt{T} \\
\sqrt{T} & -\sqrt{R}
\end{array}\right)
\end{equation}%
with $R$ and $T$ being its reflectivity and transmissivity. For a symmetric beam splitter ($R=T=0.5$), we find that $\left|\mathcal{U}_{11}\right|^2\left|\mathcal{U}_{22}\right|^2 = \left|\mathcal{U}_{12}\right|^2\left|\mathcal{U}_{21}\right|^2 = \left|\mathcal{U}_{11}\mathcal{U}_{22}\mathcal{U}^*_{12}\mathcal{U}^*_{21}\right| = 1/4$ as well as $\Phi_\mathcal{U} = \pi$, and hence the joint detection probability (\ref{pjointgeneral}) equals the well-known \eq{pjointlegero}. The cross-correlation function (\ref{g2solution}) now simplifies to
\begin{eqnarray}
\mathcal{G}^{\left(2\right)}\left(\tau\right) = \frac{1}{4\left(\tau_i + \tau_j\right)} \cdot & \left[ \exp\left(-\nicefrac{\abs{\tau} }{\tau_i}\right) + \exp\left(-\nicefrac{\abs{\tau} }{\tau_j}\right) \vphantom{\exp\left(-\gamma\left|\tau\right| -2\pi^2\Sigma^2\tau^2\right)}\right.
\nonumber \\
&\phantom{\left[\right.}\left.- 2 \exp\left(-\gamma\left|\tau\right| -2\pi^2\Sigma^2\tau^2\right)\cos{\left(2\pi\delta\nu\tau\right)}\right].
\label{g2hom}
\end{eqnarray}%
According to \eq{pcoincresult} the coincidence probability for a HOM experiment can be expressed by
\begin{equation}
p_{coinc} =\frac{1}{2} \cdot \left(1 - \frac{\mbox{Re}\left[w\left(z\right)\right]}{\sqrt{2\pi}\Sigma\left(\tau_i+\tau_j\right)} \right).
\end{equation}%
The limit for distinguishable photons (\ref{pzero}) yields $p_{coinc,0} =1/2$, which enables to write the visibility with help of \eq{visib} as
\begin{equation}
V = \frac{\mbox{Re}\left[w\left(z\right)\right]}{\sqrt{2\pi}\Sigma\left(\tau_i+\tau_j\right)}.
\label{homvisib}
\end{equation}%
Experimentally, it is common to verify the optimal spectral overlap of the photons by determining the visibility at different detunings $\delta\nu$ \cite{Patel2010, Gold2014, Giesz2015, Reindl2017}. A tuning curve $V\left(\delta\nu\right)$ of that kind, obtained by \eq{homvisib}, is illustrated in \sfig{fig2}{a} for a set of two different quantum emitters (see caption for emitter parameters). The cross-correlation function according to \eq{g2hom} is shown in \sfig{fig2}{b} and (c) in case of resonance \mbox{$\delta\nu$ = \unit[0]{GHz}} and for a detuning of \mbox{$\delta\nu$ = \unit[3]{GHz}}, respectively. It can be seen that even for a vanishing detuning a significant amount of correlated events is detected, corresponding to a visibility of just $V = $ \unit[28]{\%}. For a detuning of \unit[3]{GHz} the visibility is even further reduced to \unit[1]{\%} owing to a poor overall spectral overlap. As an additional consequence of the detuning, quantum beats become visible. Note that a widespread definition of the HOM interference visibility is given by $V = 1 - \mathcal{G}^{\left(2\right)}\left(0\right) /\mathcal{G}_0^{\left(2\right)}\left(0\right)$, i.e. a narrow time-filter is applied to the cross-correlation function instead of integrating all coincidence events according to \eq{pcoinc}. Following this definition, we find $V = $ \unit[100]{\%} for both \sfig{fig2}{b} and (c), which entirely conceals that the photons are virtually dinstinguishable. Moreover, as discussed in \cite{Legero2003}, one generally finds $\mathcal{G}^{\left(2\right)}\left(0\right) = 0$, regardless of how different both input photons are. Therefore, correlated events at $\tau= 0$ are rather a signature of a timing-jitter present in the experiment, typically caused by the photon emission itself or a slow detector response.\par
\subsection{Comparison to literature}\label{literaturecomp}
While the joint impact of PD and SD has not been theoretically investigated for remote HOM experiments so far, both effects have been considered individually. In \cite{Giesz2015} an expression for tuning curves has been derived for two independently dephasing emitters. Within the scope of the present work, the limit of vanishing inhomogeneous broadening can be obtained by re-evaluating the Fourier-transformation (\ref{fourier}) for $\alpha=0$, which leads to
\begin{equation}
V = \frac{4}{\tau_i + \tau_j}\cdot\frac{1/\tau_i + 1/\tau_j + 2\Gamma_i^* + 2\Gamma_j^*}{\left(1/\tau_i + 1/\tau_j + 2\Gamma_i^* + 2\Gamma_j^*\right)^2 + 16\pi^2\delta\nu^2}.
\label{gieszcompare}
\end{equation}%
This result is in agreement with the equation presented in \cite{Giesz2015}. Note that for two identical, but independent emitters exhibiting the radiative lifetime $\tau_r = \tau_{i,j}$, dephasing rate $\Gamma^* = \Gamma_{i,j}^*$ and $\delta\nu= 0$ this can be further simplified to the well-known expression $V = \tau_c / 2\tau_r$ \cite{Bylander2003}, with the coherence time $\tau_c$ defined by $1/\tau_c = 1/2\tau_r + \Gamma^*$.\par
The opposite limit of two independent emitters, which are inhomogeneously broadened, but do not dephase, can be described by equations (\ref{g2hom}) and (\ref{homvisib}) by setting $\Gamma_{i,j}^*=0$. In \cite{Gold2014} an equation is presented, which allows to predict the visibility for a remote HOM experiment of two resonant ($\delta\nu = 0$) emitters with identical lifetime under the influence of spectral diffusion. However, the equation does not agree with our result. For a sufficiently large inhomogeneous broadening the formalism of \cite{Gold2014} yields negative visibilities, which is an impossible outcome for HOM measurements and suggests that its derivation is erroneous. Indeed, it can be seen that the derivation is based on improperly normalized single-photon wave-functions and frequency distributions, which might cause the wrong result.\par
\Table{\label{tab1}Comparison of remote Hong-Ou-Mandel experiments to the visibility $V_{th}$ predicted by \eq{homvisib}. The experimental visibilities and accompanying theoretical values stated in the references are denoted $V_{exp}^{ref}$ and $V_{th}^{ref}$, resp. Radiative lifetime $\tau_r$, coherence time $\tau_c$, inhomogeneous linewidth $\sigma'_{max}$ and pure dephasing rate $\Gamma^*_{max}$ are given in pairs referring to emitter 1/emitter 2. All visibilities are given in percent.}
			\br
			ref.& $\tau_r$ (ps) & $\tau_c$ (ps) & $\sigma'_{max}$ (GHz) & $\Gamma^*_{max}$ (GHz) & $V_{exp}^{ref}$ & $V_{th}^{ref}$ & $V_{th}$\\
			\mr
			\cite{Gold2014} & 670/660 & 330/420 & 1.39/1.04 & 2.28/1.62 & 39 & 36$^{\rm a}$ & 28-32 \\ 
			\cite{Reindl2017} & 256/230 & 256/256 & 1.46/1.37 & 1.95/1.73 & 51 & 56 & 53-57 \\ 
			\cite{Zopf2017} & 155/187 & 153/123 & 2.46/3.53 & 3.31/5.46 & 41 & 40 & 40-44 \\ 
			\br
\end{tabular}
\item[] $^{\rm a}$ The theoretical value given in \cite{Gold2014} fits the experimental visibility better than $V_{th}$. However, note that this value was calculated based on a erroneous model (see notes in preceding paragraph).
\end{indented}
\end{table}
To independently test the validity of our formalism it is valuable to reproduce measured visibilities as stated in literature. Although a large number of remote HOM experiments on solid state emitters have been reported so far, a direct application of \eq{homvisib} is feasible only in a few cases. Many experiments are not suitable due to the non-resonant or continuous wave excitation schemes they employ \cite{Patel2010, Flagg2010, Lettow2010, Bernien2012, Sipahigil2012, Sipahigil2014}. Both add an uncertainty to the single photon emission time, which is not covered by the presented equations. Other experiments cannot be assessed, as the necessary emitter parameters are specified in an ambiguous way \cite{He2013a, Giesz2015, Thoma2017}. Moreover, no experiment can be found, for which a full set of parameters is provided, i.e. radiative lifetime, inhomogeneous linewidth and pure dephasing rate of both emitters. However, it is common to give the coherence time instead, which in turn depends on all line broadening contributions according to (cf. supplement for derivation)
\begin{equation}
\tau_c = - \frac{2\ln 2}{\pi^2}\cdot \frac{\Gamma_h}{\sigma'^2} + \sqrt{\left(\frac{2\ln 2}{\pi^2}\cdot \frac{\Gamma_h}{\sigma'^2}\right)^2 + \frac{4 \ln 2}{\pi^2 \cdot \sigma'^2}},
\label{coherence}
\end{equation}%
where the homogeneous linewidth is given by $\Gamma_h = 1/2\tau_r + \Gamma^*$. In case of three experiments \cite{Gold2014, Reindl2017, Zopf2017}, we calculate all combinations of $\sigma'$ and $\Gamma^*$ that fulfill \eq{coherence} for the given $\tau_r $ and $\tau_c$. The obtained values then lead to a range of possible visibilities $V_{th}$ according to \eq{homvisib} at $\delta\nu = 0$. All results are summarized in \tab{tab1}. The maximal values $\sigma'_{max}$ and $\Gamma^*_{max}$ given therein occur for $\Gamma^* = 0$ and $\sigma' = 0$, respectively. A comparison to the visibilities specified in the corresponding references reveals that our formalism closely reproduces the experimental values, which underlines its applicability.
\subsection{Remote HOM visibility and coherence time}\label{homandcoherence}
\begin{figure}[h!]
	\centering
	\includegraphics[width= \columnwidth]{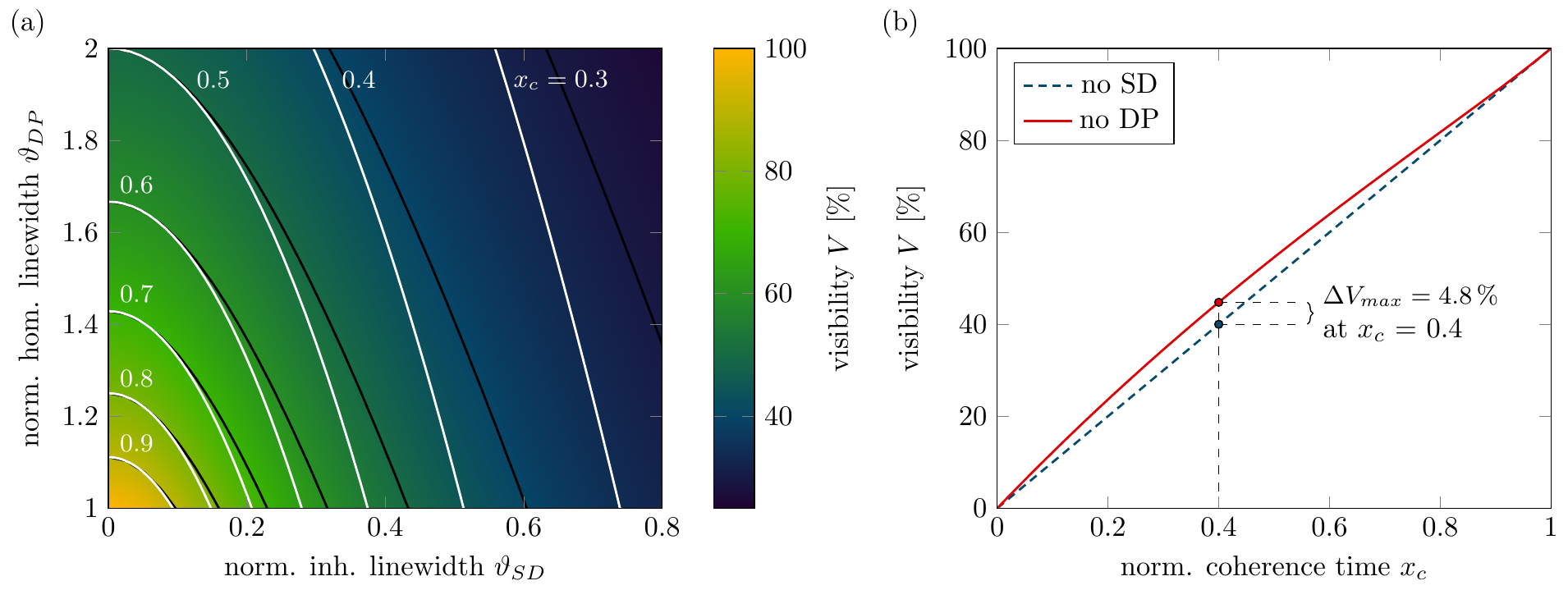}%
	\caption{Remote Hong-Ou-Mandel interference visibility for two identical emitters as a function of normalized inhomogeneous and homogeneous linewidths $\vartheta_{PD}$ and $\vartheta_{SD}$ (for definition see main text). (a) Constant levels of the normalized coherence time $x_c$ (white contour-lines) do not correspond to constant visibility (black contour-lines). (b) The limits for no spectral diffusion (blue, dashed) and no dephasing (red, solid) as a function of $x_c$ reveal the difference more clearly. At $x_c = 0.40$ both curves reach a maximum deviation of $\Delta V_{max} = 4.8\,\%$.}
	\label{fig3}
\end{figure}

In order to further quantify the simultaneous impact of PD and SD, we assume an ideal scenario of an experiment with identical emitters only, i.e. they exhibit the same lifetime \mbox{$\tau_{i,j} = \tau_r$}, same inhomogeneous broadening \mbox{$\sigma_{i,j} = \sigma \rightarrow \Sigma^2= 2\sigma^2$}, same pure dephasing rate \mbox{$\Gamma^*_{i,j} = \Gamma^*\rightarrow \gamma = 2\Gamma^* + 1/\tau_r$}, as well as no relative detuning $\delta\nu=0$. In the following, we introduce the normalized homogeneous and inhomogeneous linewidths $\vartheta_{PD} = \gamma \cdot \tau_r$ and $\vartheta_{SD} = \sigma' \cdot \tau_r$. Both parameters characterize the spectral properties of an emitter, independent of $\tau_r$. In particular, they reveal how close an emitter is to its Fourier-limit, which corresponds to $\vartheta_{PD} = 1$ and $\vartheta_{SD} = 0$. The remote HOM visibility (\ref{homvisib}) can now be written as
\begin{equation}
V = \sqrt{\frac{2 \ln 2}{\pi}} \cdot \frac{\mbox{Re}\left[w\left(z\right)\right]}{2\vartheta_{SD}} \quad \mbox{with} \quad z = \iu \sqrt{\frac{\ln 2}{2\pi^2}} \cdot \frac{\vartheta_{PD}}{\vartheta_{SD}}
\label{visibnorm}
\end{equation}%
and is depicted in \sfig{fig3}{a} as a function of $\vartheta_{PD}$ and $\vartheta_{SD}$. The bottom-left corner of the plot corresponds to Fourier-limited photons at a visibility of $V = $ \unit[100]{\%}. The additional white and black contour-lines represent levels of constant normalized coherence time $x_c = \tau_c / 2\tau_r$ following \eq{coherence} and constant visibility, respectively. The graph reveals that a high level of indinstinguishablity requires the emitters to be virtually free of PD and SD, e.g. a visibility of $V > $ \unit[90]{\%} can only be obtained for $x_c > 0.9$. In \cite{Bylander2003} it was shown that a HOM experiment with two independently dephasing emitters yields the visibility $V=x_c$, i.e. the iso-$x_c$ lines shown in \sfig{fig3}{a} are expected to run in parallel to iso-$V$ lines. However, it can be seen that they do not entirely coincide due to the additional influence of SD, which was not considered in \cite{Bylander2003}. This observation is additionally illustrated in \sfig{fig3}{b}. Here the visibility is plotted as a function of $x_c$ for the cases of no spectral diffusion $V_{noSD} = V \left(\vartheta_{PD},\vartheta_{SD}=0\right) = x_c$ and no pure dephasing $V_{noPD} = V \left(\vartheta_{PD}=1,\vartheta_{SD}\right)$. Apparently it is $V_{noPD} \geq V_{noSD}$ for all $x_c$, so that in general the visibility cannot be expressed as a function of $x_c$. However, the difference between both curves is small ($\Delta V_{max} = 4.8\,\%$ at $x_c = 0.40$) and $V_{noSD}$ serves as a reasonably accurate lower bound for all $x_c$.\par
These results further emphasize the grave impact of SD on remote TPI applications: the independent frequency-jitter of both emitters is not only non-negligible, but even corrupts the interference visibility to the same extent as PD. This conclusion might seem counterintuitive given the fact that each individual photon is Fourier-limited in the absence of PD. However, it is a direct consequence of the non-deterministic and uncorrelated relative frequency-evolution between the emitters during a long measurement run.
\subsection{Entanglement generation}
Entanglement plays a crucial role in key quantum communication and computation technologies under development \cite{Kimble2008, Nielsen2010}. Accordingly, the efficient generation of entangled quantum states is subject of numerous research activities. A well-known scheme to probabilistically entangle photons is based on a linear optical controlled-NOT (CNOT) gate \cite{Ralph2002,OBrien2003}. In the following we study this entanglement operation from the viewpoint of our formalism, as it constitutes an example of intermediate complexity that clearly demonstrates the capabilities of the presented equations. In particular, we explore limitations of the output Bell-state fidelity set by the single-photon characteristics.\par
\begin{figure}
	\centering
	\includegraphics[width= \columnwidth]{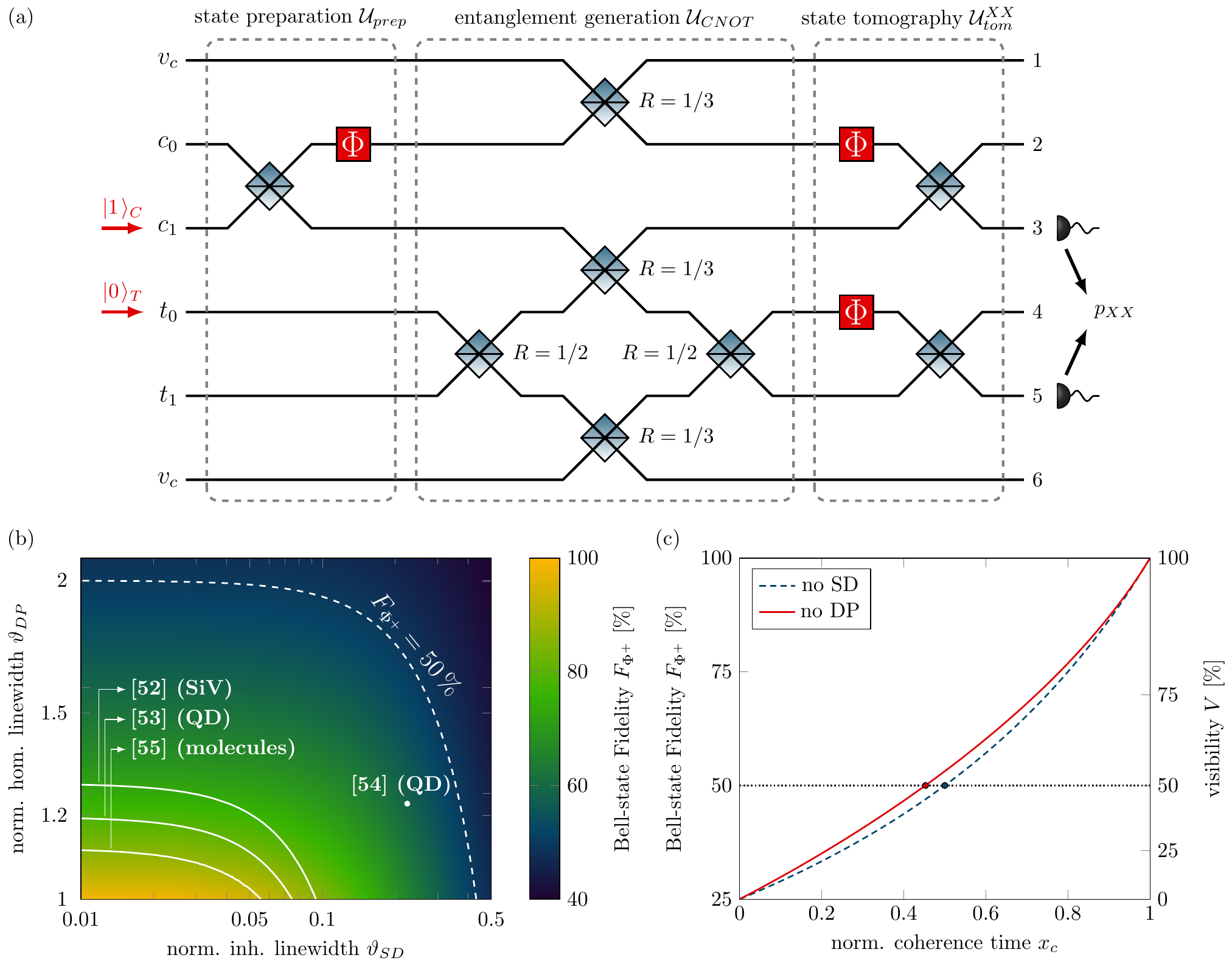}%
	\caption{Creation of the Bell-state $\left|\Phi^+\right.\rangle$ by a controlled-NOT-gate operation: (a) a control and a target photon enter the gate via $\left|1\right.\rangle_C$ and $\left|0\right.\rangle_T$. First, the control photon is prepared in a superposition state ($\mathcal{U}_{prep}$). Both photons are then entangled with help of the CNOT-gate ($\mathcal{U}_{CNOT}$). Eventually, the output state is assessed by a state tomography ($\mathcal{U}_{tom}^{XX}$ and coincidence detection). (b) Bell-state fidelity $F^{\Phi^+}$ as a function of normalized homogeneous and inhomogeneous broadening $\vartheta_{PD}$ and $\vartheta_{SD}$. The dashed white line corresponds to the limit of non-separable states at $F^{\Phi^+} = 50\,\%$. Expected fidelities for state-of-the-art emitters are included (see main text for details). (c) $F^{\Phi^+}$ as a function of the normalized coherence time $x_c$ for the cases of no spectral diffusion (blue, dashed) and no pure dephasing (red, solid).}
	\label{fig4}
\end{figure}

The basic setup of the optical gate is depicted in \sfig{fig4}{a}. It consists of 6 input and 6 output modes and is fed with 2 photons, commonly referred to as control (C) and target (T). Both photons can be considered qubits, here encoded in dual-rail representation. Accordingly, modes 2, 3 are occupied by the control qubit states $\left|0\right.\rangle_C$, $\left|1\right.\rangle_C$ and modes 4, 5 by the target qubit states $\left|0\right.\rangle_T$, $\left|1\right.\rangle_T$, respectively. The CNOT gate itself is located at the center of the allover photonic circuit. The essence of its action is to flip the state of the target photon conditioned on whether the $\left|1\right.\rangle_C$-state is occupied. The interaction between control- and target-photon is caused by TPI at the central BS connecting modes 3 and 4. Entanglement of both photons can be achieved, if the control photon is prepared in a superposition state prior to the gate. For simplicity, we only consider the creation of the Bell-state $\left|\Phi^+\right.\rangle$, which is obtained by
\begin{eqnarray}
\left|\Phi^+\right.\rangle & = \mathcal{U}_{CNOT}\left(\left|0\right.\rangle_C+\left|1\right.\rangle_C\right)\left|0\right.\rangle_T/\sqrt{2} 
\nonumber \\
& = \left(\left|0\right.\rangle_C\left|0\right.\rangle_T+\left|1\right.\rangle_C\left|1\right.\rangle_T\right)/\sqrt{2},
\end{eqnarray}%
where the gate-matrix $\mathcal{U}_{CNOT}$ was used. Our formalism is tailored for situations, in which each input photon occupies only one mode, i.e. the necessary superposition state of the control photon cannot be directly used as input. Instead, we start with a control photon in state $\left|1\right.\rangle_C$ and transform it to the desired superposition state by use of an additional BS and phase shifter. This state preparation step is represented by the matrix $\mathcal{U}_{prep}$ and can be written as $\mathcal{U}_{prep}\left|1\right.\rangle_C = \left(\left|0\right.\rangle_C+\left|1\right.\rangle_C\right)/\sqrt{2}$.\par
In order to assess the fidelity $F^{\Phi^+}$ of the output state, we emulate a state tomography following the procedure described in \cite{White2007}. Therein, the fidelity is expressed by a sum over 6 coincidence probabilities according to
\begin{equation}
F^{\Phi^+} = \left(p_{HH}+p_{VV}+p_{DD}+p_{AA}-p_{RR}-p_{LL}\right)/2.
\end{equation}
Here, the common notation for optical-polarization states was used, i.e. we identify the states $\left|H\right.\rangle = \left|0\right.\rangle$ and $\left|V\right.\rangle = \left|1\right.\rangle$ as horizontal and vertical. Accordingly, we find the superposition states diagonal $\left|D\right.\rangle=\left(\left|0\right.\rangle+\left|1\right.\rangle\right)/\sqrt{2}$, antidiagonal $\left|A\right.\rangle=\left(\left|0\right.\rangle-\left|1\right.\rangle\right)/\sqrt{2}$, right   $\left|R\right.\rangle=\left(\left|0\right.\rangle+\iu\left|1\right.\rangle\right)/\sqrt{2}$, and left $\left|L\right.\rangle=\left(\left|0\right.\rangle-\iu\left|1\right.\rangle\right)/\sqrt{2}$. The quantity $p_{XX}$ then denotes the probability to detect a coincidence between control and target photon, after projection to the state $\left|X\right.\rangle_C\left|X\right.\rangle_T$. The projection is implemented similar to the initial state preparation. We connect modes 2 and 3 as well as 4 and 5 each by a set of two linear optical components, whose transformation is expressed by the tomography-matrix $\mathcal{U}_{tom}^{XX}$. The parameters of $\mathcal{U}_{tom}^{XX}$ are chosen such that the state $\left|X\right.\rangle_C\left|X\right.\rangle_T$ is rotated to $\left|1\right.\rangle_C\left|1\right.\rangle_T$. Eventually, this enables us to calculate the desired probability $p_{XX}$ using \eq{pcoincresult} for a coincidence-measurement between the outputs $k=3$ and $l=5$. According to our discussion we use the control and target inputs $i=3$ and $j=4$ and the allover gate-matrix $\mathcal{U}_{gate} = \mathcal{U}_{tom}^{XX}\cdot\mathcal{U}_{CNOT}\cdot\mathcal{U}_{prep}$. For state preparation and tomography, we adopt the matrix elements of $\mathcal{U}_{prep}$ and $\mathcal{U}_{tom}^{XX}$ as given in \cite{James2001}. The matrix $\mathcal{U}_{CNOT}$ is constructed with the elements specified in \cite{Ralph2002}.\par
Using this procedure, the fidelity is now calculated for the normalized emitter parameters $\vartheta_{SD}$ and $\vartheta_{PD}$ in the same fashion as described in section \ref{homandcoherence}. The results for arbitrary combinations of $\vartheta_{SD}$ and $\vartheta_{PD}$ are illustrated in \sfig{fig4}{b}. The white dashed line corresponds to a Bell-state fidelity of $50\,\%$, being an important limit as it classifies any given state as separable ($<50\,\%$) or non-separable ($>50\,\%$) \cite{Terhal2000, White2007}. In \sfig{fig4}{c} both fidelity and visibility are plotted for the limiting cases of solely homogeneously and inhomogeneously broadened emitters (blue dashed and red solid curve, respectively) as a function of their normalized coherence time $x_c$. Here, it can be clearly seen that $F^{\Phi^+} = 50\,\%$ is reached at $V = 50\,\%$. Moreover, for $V \ge 50\,\%$ one finds $F^{\Phi^+} < V$, which emphasizes the high demands on single photon sources employed for entanglement schemes.\par
The results can be used to estimate $F^{\Phi^+}$, if the gate was operated with single photons from various solid-state sources. In particular, we consider the nitrogen vacancy (NV) \cite{Santori2010} and silicon vacancy (SiV) \cite{Rogers2014} defect-centers in diamond, semiconductor quantum dots (QD) \cite{Kuhlmann2015, Wei2014}, as well as single molecules in crystalline host matrices \cite{Walser2009}. The linewidths stated in literature are typically not resolved in Lorentzian and Gaussian contribution. However, similar to our approach in section \ref{literaturecomp}, we use the spectral Voigt-lineshape given in the supplement along with the known emitter lifetime $\tau_r$ to evaluate all combinations of homogeneous an inhomogeneous broadening contributions that yield the given linewidth. These combinations are translated into a function $\vartheta_{PD} \left(\vartheta_{SD}\right)$, which in turn corresponds to ranges of both visibility $V$ (\eq{visibnorm}) and fidelity $F^{\Phi^+}$ (extracted from white curves in \sfig{fig4}{b}). The results are summarized in \tab{tab2}. All systems except for the NV-center show similar fidelities in a range of \unit[73-94]{\%}, where the selected emitters are among the best specimen for each platform in terms of close-to Fourier-limited photons. Note that we assumed all operations performed by the photonic circuit to be ideal, i.e. the obtained values are only limited by the emitters' spectral properties.\par
Compared with the other presented emitter systems, the NV-center is known for a more drastic inhomogeneous broadening, typically exceeding its Fourier-limited linewdith (\unit[13]{MHz}) by an order of magnitude. In \cite{Santori2010} the emission linewidth of a single NV-center in high-quality bulk-diamond is measured by a fast scan of the excitation laser across the resonance. For a single scan, this yields a linewidth of \unit[<20]{MHz}, i.e. close to the lifetime-limit. However, repetitive scans reveal that the resonance diffuses in a range of around \unit[100]{MHz}. As this observation is a clear signature of inhomogeneous broadening, we use the single scan linewidth to obtain an upper bound for PD and find a fidelity of \unit[34-35]{\%}, mainly being limited by SD. Using a projective Bell-state measurement, it was shown that the spin-states of two remote NV-centers can be entangled with a fidelity \unit[>50]{\%} \cite{Bernien2013}, clearly exceeding the limit found here. The key to this promising outcome was to apply a frequency- and a time-filter, both of which considerably reducing the impact of SD at the cost of a reduced entanglement-generation rate.\par
\Table{\label{tab2}Remote Hong-Ou-Mandel interference visibility $V$ and Bell-state fidelity $F^{\Phi^+}$ predicted for state-of-the-art solid state single photon sources based on the lifetime and linewidth as stated in the references.}
\br
ref. & system & lifetime (ns) & linewidth (MHz) & $V$ (\%) & $F^{\Phi^+}$ (\%) \\
\mr
\cite{Santori2010} & NV & 12$^{\rm a}$ & 20/100$^{\rm b}$ & 21-23 & 34-35 \\
\cite{Rogers2014} & SiV & 1.72 & 119 & 78-91 & 73-87 \\
\cite{Kuhlmann2015} & QD & 0.85 & 270 & 84-93 & 79-91 \\
\cite{Wei2014} & QD & 0.41$^{\rm a}$ & 850$^{\rm c}$ & 62 & 59 \\
\cite{Walser2009} & molecules & 9.5 & 19 & 90-96 & 86-94 \\
\br
\end{tabular}
\item[] $^{\rm a}$ Lifetime was extracted from the given Fourier-limit.
\item[] $^{\rm b}$ Pair of values denotes {\itshape linewidth of single scan} / {\itshape spectral diffusion range}, cf. main text.
\item[] $^{\rm c}$ Voigt-linewidth based on given Lorentzian (\unit[480]{MHz}) and Gaussian (\unit[550]{MHz}) contributions.
\end{indented}
\end{table}
Eventually we like to review the QD presented in \cite{Wei2014}. The emitted photons reveal an outstanding HOM visibility of \unit[>99]{\%} for consecutive emission with a time delay of \unit[4]{ns}. Additionally, the authors show spectra that are best fit with a Lorentzian linewidth of \unit[480]{MHz} and a Gaussian linewidth of \unit[550]{MHz}. The Fourier-limited linewidth is given as \unit[390]{MHz} corresponding to a radiative lifetime of \unit[410]{ps}. Using these values, we obtain a remote HOM visibility of merely \unit[62]{\%} and a fidelity of \unit[59]{\%}. The dramatic deviation compared to the measured value is symptomatic of the often discussed time-dependency of spectral diffusion \cite{Thoma2016, Wang2016}. Within the \unit[4]{ns} passing between both emission events spectral jumps are unlikely to occur. Accordingly, the effective inhomogeneous broadening stays far below the measured linewidth, which was averaged over a much longer time. This emphasizes that the homogeneous and inhomogeneous linewidth contributions are more appropriate to assess an emitter for remote TPI applications than its HOM visibility obtained with consecutively emitted photons.
\section{Conclusion}
In the course of the present work, we derived an equation to model the joint-detection probability between any two outputs of an arbitrary linear optical gate, which was fed with two single photons and whose functionality relies on TPI. Focusing on a pair of non-identical single photons under the influence of pure dephasing and spectral diffusion as emitted by remote solid-state sources, we obtained analytical expressions to describe the time structure and allover coincidence probability of correlation measurements at the gate outputs. The model incorporates differing radiative lifetimes between both emitters, a relative detuning of their central frequencies and the influence of homogeneous and inhomogeneous broadening mechanisms. The most common application is the scattering of two photons at a single beam splitter in a HOM type interference experiment, yielding their mutual indistinguishability quantified by the interference visibility. We found that the visibility follows a Voigt line shape depending on all model parameters. Rewriting the visibility in terms of the Faddeeva-function, it is possible to numerically evaluate the expression fast and accurately. Additionally, we studied the case of a remote HOM experiment with two identical emitters being subject to spectral diffusion and pure dephasing. It turns out that the visibility cannot be displayed as a function of the photon coherence time, as it is often practiced. However, the coherence time serves as an accurate lower bound for the visibility. In particular, this implies that inhomogeneous broadening mechanisms have an equal impact on remote TPI applications compared to homogeneous linewidth contributions, even though they do not cause decoherence of individual photons.\par
To demonstrate that our formalism is applicable to more complex situations, we investigated the impact of spectral diffusion and dephasing on the Bell-state fidelity, if both photons are entangled via a CNOT gate. Besides the CNOT-operation itself, the gate contained state preparation and tomography, constituting a system with allover 11 linear optical components. The results were used to assess state-of-the-art solid-state single-photon sources as reported in literature. Even the best available emitters yield fidelities of typically below \unit[90]{\%}.\par
As it is well known that the presented limitations on remote TPI experiments are caused by PD and SD, there are various ongoing efforts aimed at reducing both effects. The most natural approach is to stabilize the emitter itself, typically by operating it under ideal conditions (low temperature, resonant excitation, Coulomb blockage) \cite{Kuhlmann2015}. Additionally, active stabilization schemes can be employed to minimize long time drifts of the emission frequency \cite{Acosta2012, Prechtel2013}. Interesting perspectives are introduced by emitters coupled to a cavity in two different regimes \cite{Grange2017}: Enhancing the spontaneous emission rate by means of the Purcell-effect broadens the natural linewidth of an emitter, thereby bringing it closer to the Fourier-limit for constant PD and SD. If in contrast an emitter is weakly coupled to a high-quality cavity, it incoherently populates the cavity mode with a single photon. Due to the weak coupling the stored photon is emitted from the cavity without any further influence from PD or SD.
\ack
The authors thank Agata Branczyk for helpful discussions in the early stages of the work, and Matthias Bock for his support in untangling the world of entanglement. This work was funded by Deutsche Forschungsgemeinschaft (DFG) (BE2306/6-1).
\section*{References}
\bibliography{TPI-limits-NJP}
\end{document}